\documentstyle[twoside,bezier,12pt]{article}

\catcode`\@=11
\long\def\@makefntext#1{ 
\protect\noindent \hbox to 3.2pt {\hskip-.9pt
$^{{\eightrm\@thefnmark}}$\hfil}#1\hfill} 

\def\thefootnote{\fnsymbol{footnote}}
 \def\@makefnmark{\hbox to 0pt{$^{\@thefnmark}$\hss}}  

\def\ps@myheadings{\let\@mkboth\@gobbletwo
\def\@oddhead{\hbox{} 
\rightmark\hfil\eightrm\thepage}
\def\@oddfoot{}\def\@evenhead{\eightrm\thepage\hfil 
\leftmark\hbox{}}\def\@evenfoot{}
\def\sectionmark##1{}\def\subsectionmark##1{}}

\oddsidemargin=\evensidemargin
\renewcommand{\thefootnote}{\fnsymbol{footnote}}

\newcounter{sectionc}\newcounter{subsectionc}\newcounter{subsubsectionc}
\renewcommand{\section}[1] {\vspace{12pt}\addtocounter{sectionc}{1}
\setcounter{subsectionc}{0}\setcounter{subsubsectionc}{0}\noindent
	{\bf\thesectionc. #1}\par\vspace{5pt}}
\renewcommand{\subsection}[1] {\vspace{12pt}\addtocounter{subsectionc}{1}
	\setcounter{subsubsectionc}{0}\noindent
	{\bf\thesectionc.\thesubsectionc. {\kern1pt \bf\it #1}}\par\vspace{5pt}}
\renewcommand{\subsubsection}[1] {\vspace{12pt}\addtocounter{subsubsectionc}{1}
	\noindent{\thesectionc.\thesubsectionc.\thesubsubsectionc.
	{\kern1pt \it #1}}\par\vspace{5pt}}

\newcommand{\textlineskip}{\baselineskip=14pt}

\def\eightcirc{
\begin{picture}(0,0)
\put(4.4,1.8){\circle{6.5}}
\end{picture}}
\def\eightcopyright{\eightcirc\kern2.7pt\hbox{\eightrm c}}

\def\abstracts#1#2#3{{
	\centering{\begin{minipage}{5in}\baselineskip=12pt\tenrm
	\centerline{ABSTRACT}
	\parindent=0pt #1\par
	\parindent=15pt #2\par
	\parindent=15pt #3
	\end{minipage} }\par}}

\renewenvironment{thebibliography}[1]			
	{
	 \begin{list}{\arabic{enumi}.}			
	{\usecounter{enumi}\setlength{\parsep}{0pt}
	 \setlength{\leftmargin 17pt}{\rightmargin 0pt}	
	 \setlength{\itemsep}{0pt} \settowidth		
	{\labelwidth}{#1.}\sloppy}}{\end{list}}	

\newcounter{itemlistc}
\newcounter{romanlistc}
\newcounter{alphlistc}
\newcounter{arabiclistc}

\newcounter{tempfigtabc}			

\def\pmb#1{\setbox0=\hbox{#1}
	\kern-.025em\copy0\kern-\wd0
	\kern.05em\copy0\kern-\wd0
	\kern-.025em\raise.0433em\box0}

\def\fnt#1#2{\footnotetext{\kern-.3em
	{$^{\mbox{\scriptsize #1}}$}{#2}}}

\def\runninghead#1#2{\pagestyle{myheadings}
\markboth{{\eightit{\quad #1}}\hfill}{\hfill{\eightit{#2\quad}}}}

 1
 1
 1

\font\eightrm=cmr8
\font\eightit=cmti8

\newcommand{\br}{\mbox{\boldmath $r$}}

\newcommand{\ba}{\begin{eqnarray}}
\newcommand{\ea}{\end{eqnarray}}
\newcommand{\be}{\begin{equation}}
\newcommand{\ee}{\end{equation}}
\newcommand{\trom}{{\rm \, tr \,}}

\def\qed{\hbox{${\vcenter{\vbox{                          
   \hrule height 0.4pt\hbox{\vrule width 0.4pt height 6pt
   \kern5pt\vrule width 0.4pt}\hrule height 0.4pt}}}$}}

\runninghead{}{}

\begin{document}
\normalsize\textlineskip
{\thispagestyle{empty}}
\setcounter{page}{1}

\renewcommand{\thefootnote}{\fnsymbol{footnote}} 
\def\bsc{{\sc a\kern-6.4pt\sc a\kern-6.4pt\sc a}}
\def\bflatex{\bf L\kern-.30em\raise.3ex\hbox{\bsc}\kern-.14em
T\kern-.1667em\lower.7ex\hbox{E}\kern-.125em X}

\centerline{\bf ELECTRIC QUADRUPOLE MOMENTS OF THE DECUPLET }
\vspace*{0.035truein}
\centerline{\bf AND}
\vspace*{0.035truein}
\centerline{\bf THE STRANGENESS CONTENT OF THE PROTON.}
\vspace{0.37truein}
\centerline{\small J. Kroll and B. Schwesinger\footnote{e-mail:
schwesinger@hrz.uni-siegen.d400.de}
}
\vspace*{0.015truein}
\centerline{\footnotesize\it FB Physik, University Siegen, Postfach 101240}
\baselineskip=12pt
\centerline{\footnotesize\it  D-57068 Siegen}
\vglue 12pt
\baselineskip=6mm
\vspace{0.37truein}
\abstracts{\noindent
In the SU3 Skyrme model the electric quadrupole moments of
$\frac{3}{2}^+$ baryons
show a strong sensitivity with respect to flavor distortions in
baryon wavefunctions.
SU3 symmetric wavefunctions lead to quadrupole moments proportional
to the charge of the baryon whereas for strongly broken flavor symmetry a
proportionality to baryonic isospin emerges.
Since the flavor distortions in the wavefunctions also determine
the strangeness content of the proton
the Skyrme model provides a link between both quantities.
\vspace{0.37truein} }{}{}

\eject

This short note is concerned with the electric quadrupole moments
$\langle B | \hat Q_{E2} | B \rangle $ of decuplet baryons $B$ where
recent quenched lattice gauge calculations\cite{LDW92} for the
non-strange members of the decuplet find them to be proportional to
their charge $\langle B | \hat Q_{E2} | B \rangle \sim
\langle B | \hat Q | B \rangle =
e \langle B | \frac{1}{2} \hat Y + \hat I_3| B \rangle $. These
calculations are in slight contradiction with
another recent prediction\cite{BSS94} from chiral perturbation theory where
the quadrupole moments are closer to a proportionality to baryonic isospin
alone, a pattern that also follows from the $SU2$-Skyrme model\cite{LDW92}.

In the Skyrme model\cite{S61,ANW83} baryons are described as
a hedgehog configuration $U_H=e^{i{\bf \tau }\cdot \hat r \chi(r)}$
performing time dependent flavor rotations $A(t)$:
\begin{eqnarray}
U({\br},t)  =  A(t) U_H({\br}) A^{\dagger}(t) \, , \quad
 A^\dagger \dot A = - \frac{i}{2} \Omega_b \lambda_b \, .
\end{eqnarray}
For such a rotating hedgehog minimal substitution of the
potential $a^0$ for a static electric field
${\bf E} = - \nabla a^0(\br)$ into the Skyrme lagrangian simply
adds the potential everywhere to the rotational velocities\cite{S93}
\begin{eqnarray}
\dot U  \rightarrow
A\left\{ [ A^\dagger \dot A,U_H] + i a^0[A^\dagger Q A ,U_H]
\right\} A^\dagger =- \frac{i}{2}\left( \Omega_b- e a^0 D_{eb}(A) \right)
A [  \lambda_b ,U_H]  A^\dagger
\end{eqnarray}
We use the standard
definitions for the D-functions in the regular representation,
$ D_{ab}(A) = \frac{1}{2} \trom \lambda_a A \lambda_b A^\dagger$,
and the abbreviation that the index $e$ stands for the linear
combination of flavors entering into the charge operator, ($
D_{ea}=D_{3a}$ for $SU2$ and $D_{ea}=
D_{3a}+\frac{1}{\sqrt{3}}D_{8a}$ for $SU(3)$ e.g.).

Rotational velocities in the Skyme model lagrangian occur in two places:
(i) quadratically in the rotational kinetic energy
\begin{eqnarray}
T_{rot}[U] \rightarrow \frac{1}{2} \int d^3r
\Theta_{ab} ({\br})( \Omega_a - e a^0 D_{ea})( \Omega_b - e a^0 D_{eb}).
\end{eqnarray}
There the density for the moments of inertia $\Theta_{ab}(\br)$
is spherical outside $SU2$-sub\-space.
Thus only the pionic inertia
$ a,b \in \left\{ 1,2,3 \right\} $  have non-spherical components.
(ii) Rotational velocities arise linearily from the anomalous
part of the action which couples the winding
number density to static electric potentials. But winding number density,
again, is spherical for hedgehogs. In contrast to the mean square radius,
i.e. the monopole moment, which receives
contributions from the pionic inertia, the kaonic inertia
$a,b \in \left\{ 4,\cdots ,7 \right\}$, the Wess-Zumino
term and further non-minimal photocouplings\cite{GL85}
\begin{eqnarray}
\Delta {\cal L} = il_9 \, a_{\mu \nu}
\,\trom Q(\nabla^\mu U \nabla^\nu U^\dagger
+ \nabla^\mu U^\dagger \nabla^\nu U)
\end{eqnarray}
the quadrupole moment originates from the rotational motion in
$SU2$-subspace alone, giving:
\begin{eqnarray}
\langle B | \hat Q_{E2} | B \rangle = \frac{e}{5}r_\pi^2
\langle B |\,  3 D_{e3} J_3 -\sum_{i=1}^3 D_{ei} J_i  \, | B \rangle
\end{eqnarray}
where
\begin{eqnarray}
 r^2_\pi = \frac{ \int d^3r r^2 \Theta_\pi(r)}{ \int d^3r \Theta_\pi(r) }
\, , \qquad
\Theta_\pi(r)=\frac{1}{3}\sum_{a=1}^3 \Theta_{aa}({\br}) \, ,
\end{eqnarray}
is the pionic contribution to - and different from - the isovectorial r.m.s.
radius of the nucleon, (see ref.\cite{SW92}).

The matrix elements of the quadrupole operator in eq(5), involve the spin
operator $J_a= -\int d^3r \Theta_\pi \Omega_a$ of the baryon
and must be calculated using the
eigenfunctions diagonalizing the rotational hamiltonian plus the SU3-symmetry
breaking terms, a method pioneered by Yabu and Ando\cite{YA88} and
refined to a "slow rotator approximation" in ref.\cite{SW91,SW92}.
Two limiting cases, however,
may be  given explicitly, the $SU3$-symmetric case and the limit where
$SU3$-symmetry breaking becomes infinite:
\begin{eqnarray}
\langle B | \hat Q_{E2} | B \rangle =
\left\{ \begin{array}{ll}
 -\frac{1}{10}\, e \, r_\pi^2 \langle B | \frac{1}{2} \hat Y
+ \hat I_3 | B \rangle & SU3-{\rm symmetric~case,} \\
{\rm ~~~}&{\rm ~~~}\\
 -\frac{4}{25}\, e \, r_\pi^2 \alpha_B
\langle B | \hat I_3  | B \rangle &
 {\rm strong~ symmetry~ breaking~ limit, }\\
{\rm ~~~} & {\rm with~}\alpha_B=\{1,\frac{5}{4},\frac{5}{3},-\}\\
{\rm ~~} & {\rm for ~}B=\{\Delta,\Sigma^*,\Xi^*,\Omega \}.
\end{array}
\right.
\end{eqnarray}
The quadrupole moments of the $\Delta$'s in the strong symmetry
breaking limit coincide with the expressions given in the $SU2$-Skyrme
model\cite{LDW92}. One can see that, as the $SU3$-symmetry breaking terms
increase the admixture of higher representations to
the pure decuplet wave functions, contributions proportional to the
hypercharge of the
baryon must become suppressed. On the other hand, a stronger mixing of
higher representations also leads to a reduction of the strangeness
content, $\langle \bar s s \rangle_B$, in the baryons. Thus we can
correlate the quadrupole moments with the strangeness content
for e.g. the proton. In the Skyrme model the latter ranges from
$\langle \bar s s \rangle_p=0$ in the strong symmetry
breaking limit to $\langle \bar s s \rangle_p=\frac{7}{30}$
for $SU3$-symmetry\cite{DN86}. The figure compiles this
information for the baryon
decuplet and one can see that for the case of the $\Delta$'s a
proportionality with respect to isospin is already reached
for empirical symmetry
breaking fixed by physical meson masses and decay constants in the
lagrangian (vertical line at $\langle \bar s s \rangle_p=.16$). With
decreasing hypercharge the quadrupole moments, however, tend to the
$SU3$-symmetric limit.
\begin{figure}[h]
\begin{center} \setlength{\unitlength}{.2cm}
\begin{picture}(60.,28.)(-1.,-16.) \thinlines
\put(0.,-15.){\vector(0,1){30.}}
\put(1.,14.){$Q_{E2}/r_\pi^2$}
\put(-3.,-10.){-.2}
\put(-3.,-5.){-.1}
\put(-3., 5.){0.1}
\put(-3., 10.){0.2}
\put(0.,0.){\vector(1,0){60.}}
\put(56.,-2.){$\langle {\bar s} s \rangle _p $}
\put(20.,-2.){.1}
\put(40.,-2.){.2}
\thinlines
\put(33.33,-12.5){\line(0,1){25.}}
\multiput(33.,-10.)(0,5.0){5}{\line(1,0){0.33}}

\multiput(-.3,-10.)(0,5.0){5}{\line(1,0){0.3}}
\multiput(20.,-.3)(20.0,0.){2}{\line(0,1){0.3}}
\put(48.,-9.7){$\Delta^{++}$}
\put(48.,-4.7){$\Delta^{+}$}
\put(48.,.3){$\Delta^{0}$}
\put(48.,5.3){$\Delta^{-}$}
\thicklines
 \bezier{100}( 46.67,  5.00)( 37.79,  8.20)( 33.33,  9.17)
 \bezier{100}( 46.67,  0.00)( 37.79,  2.08)( 33.33,  2.69)
 \bezier{100}( 46.67, -5.00)( 37.79, -4.05)( 33.33, -3.80)
 \bezier{100}( 46.67,-10.00)( 37.79,-10.17)( 33.33,-10.29)
 \bezier{100}( 33.33,  9.17)( 29.13,  9.79)( 26.03, 10.14)
 \bezier{100}( 33.33,  2.69)( 29.13,  3.05)( 26.03,  3.24)
 \bezier{100}( 33.33, -3.80)( 29.13, -3.69)( 26.03, -3.65)
 \bezier{100}( 33.33,-10.29)( 29.13,-10.43)( 26.03,-10.55)
 \bezier{100}( 26.03, 10.14)( 23.68, 10.36)( 21.83, 10.52)
 \bezier{100}( 26.03,  3.24)( 23.68,  3.36)( 21.83,  3.44)
 \bezier{100}( 26.03, -3.65)( 23.68, -3.65)( 21.83, -3.65)
 \bezier{100}( 26.03,-10.55)( 23.68,-10.65)( 21.83,-10.74)
 \bezier{100}( 21.83, 10.52)( 19.12, 10.74)( 17.20, 10.89)
 \bezier{100}( 21.83,  3.44)( 19.12,  3.54)( 17.20,  3.60)
 \bezier{100}( 21.83, -3.65)( 19.12, -3.67)( 17.20, -3.69)
 \bezier{100}( 21.83,-10.74)( 19.12,-10.88)( 17.20,-10.98)
 \bezier{100}( 17.20, 10.89)( 12.27, 11.23)(  5.52, 11.66)
 \bezier{100}( 17.20,  3.60)( 12.27,  3.73)(  5.52,  3.89)
 \bezier{100}( 17.20, -3.69)( 12.27, -3.76)(  5.52, -3.89)
 \bezier{100}( 17.20,-10.98)( 12.27,-11.26)(  5.52,-11.67)
 \bezier{100}(  5.52, 11.66)(  3.91, 11.76)(  0.00, 12.00)
 \bezier{100}(  5.52,  3.89)(  3.91,  3.92)(  0.00,  4.00)
 \bezier{100}(  5.52, -3.89)(  3.91, -3.92)(  0.00, -4.00)
 \bezier{100}(  5.52,-11.67)(  3.91,-11.76)(  0.00,-12.00)
\end{picture} \end{center}
\end{figure}


\begin{figure}[h]
\begin{center} \setlength{\unitlength}{.2cm}
\begin{picture}(60.,24.)(-1.,-12.) \thinlines
\put(0.,-11.){\vector(0,1){26.}}
\put(1.,14.){$Q_{E2}/r_\pi^2$}
\put(-3.,-10.){-.2}
\put(-3.,-5.){-.1}
\put(-3., 5.){0.1}
\put(-3., 10.){0.2}
\put(0.,0.){\vector(1,0){60.}}
\put(56.,-2.){$\langle {\bar s} s \rangle _p $}
\put(20.,-2.){.1}
\put(40.,-2.){.2}
\thinlines
\put(33.33,-9.5){\line(0,1){22.}}
\multiput(33.,-5.)(0,5.0){4}{\line(1,0){0.33}}

\multiput(-.3,-10.)(0,5.0){5}{\line(1,0){0.3}}
\multiput(20.,-.3)(20.0,0.){2}{\line(0,1){0.3}}
\put(48.,-4.7){$\Sigma^{*+}$}
\put(48.,.3){$\Sigma^{*0}$}
\put(48.,5.3){$\Sigma^{*-}$}
\thicklines
 \bezier{100}( 46.67,  5.00)( 37.79,  7.25)( 33.33,  8.09)
 \bezier{100}( 46.67,  0.00)( 37.79,  0.76)( 33.33,  0.99)
 \bezier{100}( 46.67, -5.00)( 37.79, -5.73)( 33.33, -6.11)
 \bezier{100}( 33.33,  8.09)( 29.13,  8.64)( 26.03,  8.93)
 \bezier{100}( 33.33,  0.99)( 29.13,  1.08)( 26.03,  1.07)
 \bezier{100}( 33.33, -6.11)( 29.13, -6.48)( 26.03, -6.79)
 \bezier{100}( 26.03,  8.93)( 23.68,  9.10)( 21.83,  9.22)
 \bezier{100}( 26.03,  1.07)( 23.68,  1.03)( 21.83,  0.99)
 \bezier{100}( 26.03, -6.79)( 23.68, -7.04)( 21.83, -7.25)
 \bezier{100}( 21.83,  9.22)( 19.12,  9.36)( 17.20,  9.45)
 \bezier{100}( 21.83,  0.99)( 19.12,  0.90)( 17.20,  0.82)
 \bezier{100}( 21.83, -7.25)( 19.12, -7.57)( 17.20, -7.81)
 \bezier{100}( 17.20,  9.45)( 12.27,  9.64)(  5.52,  9.85)
 \bezier{100}( 17.20,  0.82)( 12.27,  0.60)(  5.52,  0.28)
 \bezier{100}( 17.20, -7.81)( 12.27, -8.43)(  5.52, -9.30)
 \bezier{100}(  5.52,  9.85)(  3.91,  9.90)(  0.00, 10.00)
 \bezier{100}(  5.52,  0.28)(  3.91,  0.20)(  0.00,  0.00)
 \bezier{100}(  5.52, -9.30)(  3.91, -9.51)(  0.00,-10.00)
\end{picture} \end{center}
\end{figure}


\begin{figure}[h]
\begin{center} \setlength{\unitlength}{.2cm}
\begin{picture}(60.,20.)(-1.,-8.) \thinlines
\put(0.,-8.){\vector(0,1){23.}}
\put(1.,14.){$Q_{E2}/r_\pi^2$}
\put(-3.,-5.){-.1}
\put(-3., 5.){0.1}
\put(-3., 10.){0.2}
\put(0.,0.){\vector(1,0){60.}}
\put(56.,-2.){$\langle {\bar s} s \rangle _p $}
\put(20.,-2.){.1}
\put(40.,-2.){.2}
\thinlines
\put(33.33,-5.5){\line(0,1){17.}}
\multiput(33.,-5.)(0,5.0){4}{\line(1,0){0.33}}

\multiput(-.3,-5.)(0,5.0){4}{\line(1,0){0.3}}
\multiput(20.,-.3)(20.0,0.){2}{\line(0,1){0.3}}
\put(48.,.3){$\Xi^{*0}$}
\put(48.,5.3){$\Xi^{*-}$}
\thicklines
 \bezier{100}( 46.67,  5.00)( 37.79,  6.12)( 33.33,  6.62)
 \bezier{100}( 46.67,  0.00)( 37.79, -0.77)( 33.33, -1.23)
 \bezier{100}( 33.33,  6.62)( 29.13,  6.96)( 26.03,  7.11)
 \bezier{100}( 33.33, -1.23)( 29.13, -1.73)( 26.03, -2.16)
 \bezier{100}( 26.03,  7.11)( 23.68,  7.17)( 21.83,  7.19)
 \bezier{100}( 26.03, -2.16)( 23.68, -2.53)( 21.83, -2.84)
 \bezier{100}( 21.83,  7.19)( 19.12,  7.17)( 17.20,  7.14)
 \bezier{100}( 21.83, -2.84)( 19.12, -3.33)( 17.20, -3.68)
 \bezier{100}( 17.20,  7.14)( 12.27,  7.03)(  5.52,  6.84)
 \bezier{100}( 17.20, -3.68)( 12.27, -4.57)(  5.52, -5.74)
 \bezier{100}(  5.52,  6.84)(  3.91,  6.79)(  0.00,  6.67)
 \bezier{100}(  5.52, -5.74)(  3.91, -6.01)(  0.00, -6.67)
\end{picture} \end{center}
\end{figure}


\begin{figure}[h]
\begin{center} \setlength{\unitlength}{.2cm}
\begin{picture}(60.,15.)(-1.,-3.) \thinlines
\put(0.,-2.){\vector(0,1){17.}}
\put(1.,14.){$Q_{E2}/r_\pi^2$}
\put(-3., 5.){0.1}
\put(-3., 10.){0.2}
\put(0.,0.){\vector(1,0){60.}}
\put(56.,-2.){$\langle {\bar s} s \rangle _p $}
\put(20.,-2.){.1}
\put(40.,-2.){.2}
\thinlines
\put(33.33,-1.){\line(0,1){13.}}
\multiput(33.,5.)(0,5.0){2}{\line(1,0){0.33}}

\multiput(-.3,0.)(0,5.0){3}{\line(1,0){0.3}}
\multiput(20.,-.3)(20.0,0.){2}{\line(0,1){0.3}}
\put(48.,5.3){$\Omega^{-}$}
\thicklines
 \bezier{100}( 46.67,  5.00)( 37.79,  4.82)( 33.33,  4.65)
 \bezier{100}( 33.33,  4.65)( 29.13,  4.41)( 26.03,  4.11)
 \bezier{100}( 26.03,  4.11)( 23.68,  3.80)( 21.83,  3.49)
 \bezier{100}( 21.83,  3.49)( 19.12,  3.01)( 17.20,  2.67)
 \bezier{100}( 17.20,  2.67)( 12.27,  1.86)(  5.52,  0.83)
 \bezier{100}(  5.52,  0.83)(  3.91,  0.59)(  0.00,  0.00)
\put(0.,-8.){\parbox{13.cm}
{\footnotesize Figure 1: Quadrupole moments versus the
strangeness content of the proton. The functions have been obtained in
rigid rotator approximation, ref.\cite{SW91}, with the parameters quoted
there. The strangeness content is varied through a change of the
kaon mass $m_K$.
The vertical line indicates the position where $m_K=495$MeV.}}
\end{picture} \end{center}
\end{figure}
\clearpage
In plotting the quadrupole moments versus the strangeness content of
the proton we hope to have removed some model dependence
out of our statements, which are contained, for example, in assumptions on the
exact form of higher order terms in the effective lagrangian.
Nevertheless, we would like to terminate this short note by presenting
(model dependent) numbers for these moments in table 1. as they
follow from the slow rotator approach (case SK4 in ref.\cite{SW92}) to the
$SU3$-rotational motion of the soliton: \\[1.cm]
\begin{tabbing}
123456\=12345678\=12345678901234567890123456\=123456\=12345678
\=1234567890\=\kill
B\> SRA \> CPT\>B\> SRA \> CPT\\
\\
$\Delta^{++}$\>-0.87\>-0.8$\pm$
0.5\>$\Sigma^{*+}$\>-0.42\>-0.7$\pm$ 0.3\\
$\Delta^{+}$\>-0.31\>-0.3$\pm$
0.2\>$\Sigma^{*0}$\>+0.05\>-0.13$\pm$0.07\\
$\Delta^{0}$\>+0.24\>+0.12$\pm$ 0.05\>$\Sigma^{*-}$\>+0.52\>+0.4
$\pm$0.2\\
$\Delta^{-}$\>+0.80\>+0.6$\pm$ 0.3\>\>\>\\
\>\>\>$\Xi^{*0}$\>-0.07\>-0.35$\pm$ 0.2\\
$\Omega^{-}$\>+0.24\>+0.09$\pm$ 0.05\>$\Xi^{*-}$\>+0.35\>+0.2$\pm$ 0.1
\end{tabbing}
{\small Table 1. Quadrupole moments of the baryon decuplet in units
${10^{-1} {\rm e\cdot fm}^2}$ in slow rotator
approximation (SRA) compared to chiral perturbation theory\cite{BSS94} (CPT).}
\\[.75cm]
For the non-strange $\Delta$'s chiral perturbation theory and the
Skyrme model, both approaches, find a pattern proportional
to isospin which in the case of chiral perturbation theory persists
also for the strange members of the decuplet, whereas the Skyrme model
moves closer to charge proportionality. As far as the magnitude of the
moments is concerned, there seems to be mutual agreement, but at least
in the case of the Skyrme model calculation there
is a caveat : the
isovector radius $\langle r^2 \rangle_V = \langle r^2 \rangle_p-\langle
r^2 \rangle_n$ comes out too small
( $\langle r^2 \rangle_V=.49 {\rm fm}^2$ for case SK4 in ref.\cite{SW92} ).
Thus, the ratio of quadrupole moment to isovector radius is
rather high in the Skyrme model as was correctly noticed in ref.\cite{LDW92}.
In the slow rotator approximation it is roughly a
factor of two higher than the corresponding ratio of the lattice gauge
calculation\cite{LDW92}.

In conclusion we have shown that three different approaches to the
electric quadrupole moments of decuplet baryons: a quenched lattice gauge
calculation, chiral perturbation theory and the Skyrme model, lead to three
slightly different predictions which apparently differ in the amount of
$SU3$ symmetry breaking in the decuplet states: the lattice gauge
results are closest to $SU3$ symmetry whereas chiral perturbation
theory is closest to the strong symmetry breaking limit. The $SU3$
Skyrme model, finally, allows a smooth interpolation between the two
limits as a function of the strangeness content of the proton and
predicts a pattern intermediate between those mentioned, if empirical
$SU3$ symmetry breaking is employed.

\vspace{5.truein}

\clearpage

\begin{thebibliography}{10}


\bibitem{LDW92}
D. B. Leinweber, T. Draper and R. M. Woloshyn, Phys. Rev. \underline{D46}
(1992) 3067

\bibitem{BSS94}
M. N. Butler, M. J. Savage and R. P. Springer, Phys. Rev.
\underline{D49} (1994) 3459

\bibitem{S61}
T. H. R. Skyrme, Proc. Roy. Soc. \underline{A260} (1961) 127

\bibitem{ANW83}
G. S. Adkins, C. R. Nappi and E. Witten, Nucl. Phys. \underline{B228}
(1983) 552

\bibitem{S93}
B. Schwesinger, Phys. Lett. \underline{B298} (1993) 17

\bibitem{GL85}
J. Gasser and H. Leutwyler, Nucl. Phys. \underline{B250} (1985) 465

\bibitem{SW92}
B. Schwesinger and H. Weigel, Nucl. Phys. \underline{A450} (1992) 461

\bibitem{YA88}
H. Yabu and K. Ando, Nucl. Phys. \underline{B301} (1988) 601

\bibitem{SW91}
B. Schwesinger and H. Weigel, Phys. Lett. \underline{B267} (1991) 438

\bibitem{DN86}
J. F. Donoghue and C. R. Nappi, Phys. Letts. \underline{B168} (1986) 105

\end{thebibliography}
\end{document}